\documentclass[12pt]{iopart}
\usepackage{graphicx}
\usepackage[normalem]{ulem}\usepackage{soul}
\usepackage{iopams}
\usepackage{color}
\newcommand{\be}{\begin{equation}}
\newcommand{\ee}{\end{equation}}
\newcommand{\PT}{$\mathcal{PT}$}

\begin{document}

\title[Jamming anomaly in $\mathcal{PT}$-symmetric systems]{Jamming anomaly in $\mathcal{PT}$-symmetric  systems}

\author{I V Barashenkov}

\address{
Department of Mathematics, University of Cape Town,  Rondebosch 7701
and
National Institute for Theoretical Physics, Western Cape, South Africa
 }
\ead{Igor.Barashenkov@uct.ac.za}

\author{D  A Zezyulin}
\address{Centro de F\'isica Te\'orica e Computacional and Departamento de F\'isica,
Faculdade de Ci\^encias da Universidade de Lisboa, Campo Grande, Edif\'icio C8, Lisboa  P-1749-016, Portugal}
\ead{dzezyulin@fc.ul.pt}

\author{V V Konotop}
\address{Centro de F\'isica Te\'orica e Computacional and Departamento de F\'isica,
 Faculdade de Ci\^encias da Universidade de Lisboa, Campo Grande, Edif\'icio C8, Lisboa  P-1749-016, Portugal}
 \ead{vvkonotop@fc.ul.pt}
\vspace{10pt}
\begin{indented}
\item[]March 2016
\end{indented}

\begin{abstract}

The Schr\"odinger equation with a
 $\mathcal{PT}$-symmetric
potential  is used
to model
 an optical structure consisting  of an element with  gain
coupled to an element with loss.
At low gain-loss amplitudes $\gamma$,  raising the amplitude results in
the energy
flux from the active to the leaky element being boosted.
We study the  anomalous behaviour occurring for larger $\gamma$,
 where the increase of the amplitude produces a drop of the
flux across the gain-loss interface.
We show that this {\it jamming anomaly\/}  is either a precursor of the exceptional point, where
two real eigenvalues coalesce and acquire imaginary parts, or precedes the
eigenvalue's immersion in the continuous spectrum.

\end{abstract}

%
%
%
%
%

\section{Introduction}

We consider a spatially extended system consisting of two coupled elements, where the energy is gained in
the first element  and  dissipated in the second one.
To make stationary processes possible,  the structure is designed so that the gain exactly balances the loss.
The energy is assumed to be gained and dissipated locally, that is,
 there are no flows to or from infinity.
Mathematically, this system is modelled by a parity-time (\PT) symmetric  Schr\"odinger equation,  linear or non-linear, with a two-centre localised complex potential
and vanishing boundary conditions.

This  bicentric structure is prototypical for a variety of physical settings, such as the
electromagnetic field between  the active and passive coupled  parallel
waveguides~\cite{waveguide,Kivshar_review} and
\textcolor{black}{microtoroid resonators \cite{PT_breaking}},
 or
 the pumped and lossy atomic cells~\cite{Muga}.
 A similar gain-loss dimer is formed by the
  Bose-Einstein condensate trapped in two coupled wells, with one well leaking and the other one being loaded with atoms  \cite{Cartarius}.
  One more context is provided by the quasi-one-dimensional polaritonic waveguide.
Here, the bicentric  gain-loss structure can be modelled on  the transistor switch realised experimentally in \cite{polariton}.
  The  localized injection of polaritons at one point and enhanced absorption at another point of the waveguide
  can be exploited as a means of the current  control.
   \textcolor{black}{A  \PT-symmetric system consisting of  two coupled  mechanical resonators, a  damped nonlinear and
 a driven linear  one, also deserves to be mentioned \cite{Mech}.
  The system was proposed as a concept for
 an on-chip microscale phonon diode. (More \PT-symmetric bicentric structures are discussed in
 a recent review \cite{KYZ} where one can find all the pertinent references.)
 }

Stationary processes in the bicentric structure are sustained by the energy flux
  from the active to the passive element.
 We study the control of the flux --- and therefore, control
 of the total gain and loss rate in the corresponding elements ---
 by means of the variations of  the gain-loss amplitude  of the system.
 Intuitively, the increase of this gain and loss factor
 is expected to intensify the flux. The aim of this paper is to show that
 the
 anomalous behaviour is also possible; namely,
  the flux  may drop as a result of the gain-loss amplitude  increase.
  For reasons explained below,
   we are referring to this phenomenon as the {\it jamming anomaly.}


The outline of the paper is as follows.
After formulating the anomaly mathematically
(section \ref{Formulation}), we exemplify the general concept with two exactly solvable models.
First, we demonstrate the anomalous behaviour 
in the linear Schr\"odinger equation with the \PT-symmetric
double-delta well  potential (section \ref{double_lin})
and in its  non-linear counterpart (section \ref{double_nonlin}).
To show that the double well is not imperative for the anomaly
 to occur, we then consider
a potential with a single  well (the \PT-symmetric Scarff potential, section \ref{Sca}).

\textcolor{black}{
Both potentials considered reveal jamming near their exceptional points or just before the
corresponding eigenvalue immerses in the
continuous spectrum.
In  section \ref{exceptional}, we argue that {\it any\/}  \PT-symmetric potential
with an exceptional point in its spectrum
should exhibit the flux anomaly in the gain-loss parameter region adjacent to that point.
To test whether the proximity to an exceptional point
(or to the edge of the continuum) is not only sufficient but necessary for the anomaly to occur,
we examine an exactly-solvable example of a potential  with an entirely discrete spectrum free from  exceptional points
(section \ref{harmonic}). The corresponding flux shows no anomalous behaviour.}

The conclusions of this study are summarised in section \ref{conclusions}.


\section{Gain, loss and jamming}
\label{Formulation}

\subsection{Flux across the gain-loss interface}

The system is described by  the nonlinear Schr\"odinger equation,
\be
i \Psi_t + \Psi_{xx} - V(x)
 \Psi + g\Psi|\Psi|^2 = 0,
\label{A0}
\ee
or its linear counterpart defined by setting $g=0$ in (\ref{A0}).
Here $\Psi$
is either a dimensionless amplitude of the electric field or an order parameter --- depending on whether one aims for the
optical or boson-condensate interpretation of equation \eref{A0}. In the context of
atomic and polaritonic condensates,
$t$ denotes a properly scaled time.  In the
  paraxial optics application, $t$ measures the propagation distance  while $x$ is the coordinate transversal to the beam.

In equation  \eref{A0},   $V(x)$ is  the  $\mathcal{PT}$-symmetric potential:
\[
V(x)= U(x) + i W (x),
\]
with $U(-x)=U(x)$ and $W(-x)=-W(x)$.
\textcolor{black}{A product of the
 nonhermitian quantum mechanics \cite{Bender1,Bender2}, the
\PT-symmetric potentials are becoming increasingly relevant
 in optics
  and other applied disciplines. (See e.g. the recent reviews \cite{Kivshar_review,KYZ}.)}

\begin{figure}[t]
\begin{center}
\includegraphics[width=0.6\linewidth]{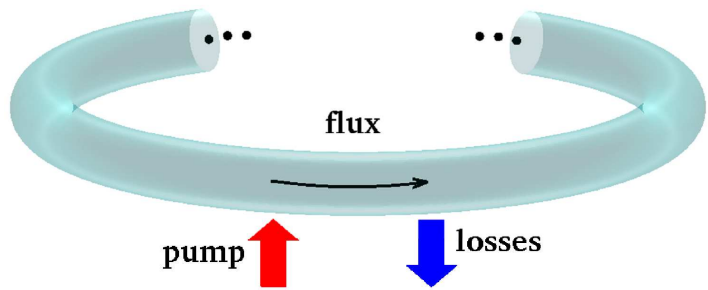}\end{center}
\caption{\label{ring}
The bicentric gain-loss system: the energy is pumped into 
a local site and is dissipated in its mirror-image site.
The gain and loss sites are assumed to be centred close enough to each other so that
 the system can be thought of having the geometry of an infinite line
 --- even if it is ring-shaped.
}
\end{figure}

For simplicity of the analysis, we assume that $x$ lies on the infinite line, $-\infty<x<\infty$.
The right half of the $x$-line is characterised by loss
and the left half  by gain:
\[
W(x)> 0 \ \mbox{if} \ x<0;
\qquad
W(x)<0 \ \mbox{if} \ x >0.
\]
The infinite line can serve as an approximation for a ring-shaped structure
(see Fig.\ref{ring}),
provided $U(x)$ and $W(x)$ are rapidly decaying functions with the decay ranges
 much shorter than the length of the ring.
We consider localised solutions of  equation \eref{A0} --- that is, solutions with the asymptotic behaviour
$\Psi(x,t) \to 0$ as $x \to \pm \infty$.

The integrals $\int_{-\infty}^0|\Psi|^2 dx$
and
  $\int_0^\infty |\Psi|^2 dx$
 give the beam power
  (or  the number of particles) captured in the regions
with gain and loss, respectively.
The power (the number)   in the dissipative region varies according to
\be
 \frac{d}{dt} \int_0^\infty |\Psi|^2 dx=   J(t)   + 2  \int_0^\infty |\Psi|^2 \, W(x)   dx,
\label{A00}
\ee
where we have defined the flux across the gain-loss interface:
\be
\left.  \phantom{\frac12}   J= i (\Psi_x^* \Psi- \Psi^* \Psi_x) \right|_{x=0}.
 \label{A01}
\ee

\textcolor{black}{
Equation \eref{A00} implies that two sources of the power variation
 in the dissipative domain, are (a)  the energy flux from the  region with gain
 and  (b) the losses suffered  between $x=0$ and $\infty$.
 Likewise, the variation of the power between $x=0$ and $-\infty$,
 is given by the gain in that region less the flux through the origin.}

In this paper, we focus on stationary regimes, $\Psi(x,t)=\psi(x)e^{i \kappa^2 t}$. Here the real
$\kappa^2$ represents the propagation constant in optics and chemical potential in the context
of boson condensates.
The spatial part of the separable solution $\Psi$ satisfies
\be
-\psi_{xx} + V(x)
 \psi - g\psi|\psi|^2 = -\kappa^2 \psi.
\label{A1}
\ee
The boundary conditions
\be
\psi(x) \to 0 \quad \mbox{as} \ x \to \pm \infty
\label{BC}
\ee
require that $\kappa^2$ be taken positive.


In the stationary regime, the power density $|\Psi|^2$ and the flux \eref{A01} are time-independent:
$|\Psi|^2=|\psi(x)|^2$ and
\be
\left. \phantom{\frac12}  J= i (\psi_x^* \psi- \psi^* \psi_x) \right|_{x=0}.
 \label{D20}
 \ee
 In this case equation   \eref{A00} gives
\be
 J  =-2  \int_0^\infty |\psi|^2  \,W(x)   \, dx>0.
\label{A2}
\ee

Assume that the  rate of gain and loss is controlled by a single parameter $\gamma>0$;
for instance,  let
\[
W= \gamma \mathcal{W} (x),
\]
where $\mathcal{W} (x)$ is $\gamma$-independent.
We will be referring to $\gamma$ as the gain-loss amplitude.

\textcolor{black}{
In the linear case ($g=0$), the  constant  $-\kappa^2$ arises as an eigenvalue of the Schr\"odinger operator \eref{A1}.
For the given $\gamma$, there is a \textcolor{black}{sequence}  of eigenvalues $-\kappa_n^2$, $n=0, 1, 2, ...$.
When $\gamma=0$, the operator \eref{A1} is hermitian and all eigenvalues are real.
The eigenvalues remain on the real line even when $\gamma$ is made small nonzero; this is a fascinating consequence of the \PT \, symmetry \cite{Bender1,Bender2}.
As $\gamma$ is increased further, then, at   some  point $\gamma_c$ (commonly referred to as the {\it exceptional point} \cite{parab1,Heiss}),
one or more  pairs of eigenvalues may merge and become complex. In the region above the exceptional point, the
\PT \, symmetry is said to be spontaneously broken \cite{Bender1,Bender2,PT_breaking}. }

\textcolor{black}{
The present paper deals with the unbroken region, $\gamma<\gamma_c$.
Here all eigenvalues $-\kappa_n^2$ are real, and each of  the corresponding eigenfunctions $\psi_n(x)$ can be brought to
the \PT-symmetric form by a  constant phase shift \cite{Bender2,BQZ}.
That is to say, there is no loss of generality in assuming that
\be
\psi_n(-x)= \psi_n^*(x).
\label{pspt}
\ee
This will be our routine assumption in what follows.
}

\textcolor{black}{
The question that we concern ourselves with,
is how the interfacial flux $J$ associated with the normalised eigenfunction $\psi_n(x)$,
responds to the variation of  $\gamma$.
}

A similar question can be posed for the nonlinear Schr\"odinger equation.   
Unlike a generic dissipative system,
a typical \PT-symmetric potential supports a one-parameter family of localised modes
for every fixed $\gamma$ in some interval  \cite{Cartarius,KYZ,pt-potential,parab3,coupled_NLS}.
The total power carried by
the localised mode,
 \[
P=  \int_{-\infty}^\infty |\psi|^2 dx,
\]
  is a function of
   $\gamma$
   and
  $\kappa$,
  where $\kappa$ is the parameter of the family.
  Setting the power to a certain fixed value, e.g. setting $P(\gamma, \kappa)=1$,  makes  $\kappa$ an (implicit)
function of $\gamma$: $\kappa=\kappa_P(\gamma)$. Our aim is to find out how the flux $J$ associated with the nonlinear mode with the parameter
$\kappa=\kappa_P(\gamma)$ changes as $\gamma$ is varied.

\subsection{Toy model: the \PT-symmetric dimer}
\label{toy}

At  first glance, the increase of the gain-loss amplitude $\gamma$
should boost the flux:
the greater is the  power growth rate  in the active region and the faster is the
power loss in the dissipative domain, the more energy flows across the gain-loss interface.

To illustrate this intuitively appealing idea,
assume that  the active  and lossy elements  of our $\mathcal{PT}$-symmetric doublet are point-like objects
--- that is,  they do not have any internal structure.
 In this case the $\mathcal{PT}$-symmetric
 system can be described by a two-component vector equation known as the nonlinear Schr\"odinger dimer \cite{waveguide,Kivshar_review,PT_dimer}:
 \begin{eqnarray*}
i {\dot \Psi_1}  + \Psi_2+ g|\Psi_1|^2 \Psi_1   & =   \phantom{-}   i \gamma \Psi_1,  \\   
i {\dot \Psi_2}   +\Psi_1+g |\Psi_2|^2 \Psi_2   &  =  -   i\gamma \Psi_2.
\label{standard}
\end{eqnarray*}
Here $\Psi_1$ and $\Psi_2$ are complex amplitudes of  the active and
lossy modes, respectively.
The powers carried by the two modes are $|\Psi_1|^2$ and $|\Psi_2|^2$,
and
the interfacial flux is given by
\[
J=i(\Psi_1\Psi_2^*-\Psi_2\Psi_1^*).
\]

Stationary solutions
of the dimer  have the form  $\Psi_1(t)=e^{i \kappa^2 t} \psi_1$ and  $\Psi_2(t)=e^{i \kappa^2 t} \psi_2$,
where $\psi_1$ and $\psi_2$ are complex coefficients determined by initial conditions,
and $\kappa^2$ is a real propagation constant.
The stationary mode powers $P_1= |\psi_1|^2$ and $P_2=|\psi_2|^2$
are equal:
 $P_{1,2}= P$.
The  corresponding flux is
\be
J= 2 \gamma P.
\label{AB1}
\ee
The modes' common power is
 related to $\kappa$ and $\gamma$ by one of the following two equations:
\be
 P=\frac{1}{g} \left(
 \kappa^2 \pm \sqrt{1-\gamma^2} \right).
 \label{F60}
 \ee

In order  to  determine the effect of the  gain-loss amplitude variation on the flux \eref{AB1}, we
consider the initial conditions $\psi_{1,2}$ with the
$\gamma$-independent mode powers: $P_1=P_2$ ($=P$).
In the linear case ($g=0$) this simply means that  the  vector
 $( \psi_1, \psi_2 )$ is normalised; for example,  we can let  $P=1$.
 In the nonlinear situation ($g \neq 0$) we can also choose $P=1$;
this choice selects  one
value of the propagation constant out of the family in \eref{F60}:
 $\kappa^2= g \mp \sqrt{1-\gamma^2}$.

  Differentiating \eref{AB1} with respect to $\gamma$  and keeping in mind that $dP/ d\gamma=0$,
  gives
\[
\frac{dJ}{d \gamma} = 2 P.
\]
This is always positive --- in agreement with our intuition.


\subsection{Jamming anomaly in the \PT-symmetric Schr\"odinger equation}

Returning to  our spatially extended system,  equation \eref{A2} yields
\be
\frac{dJ}{d \gamma} = -2 \int_0^\infty  |\psi|^2 \,
 \mathcal{W}  \, dx- 2 \gamma \int_0^\infty  \frac{\partial |\psi|^2}{\partial \gamma} \mathcal{W}  \,
 dx.
\label{st}
\ee
The first term in the right-hand side of \eref{st} is positive.  The second term  can be negative,
but if $\gamma$ is small, the sum shall nevertheless be positive,
$dJ/d \gamma>0$, --- as in the structureless example of section \ref{toy}.

Note that the initial conditions giving rise to stationary solutions of
 \eref{A0} cannot be chosen arbitrarily --- in particular, one can
 generally not choose
 $\partial |\psi|^2/ \partial \gamma=0$.
The dependence of $\psi(x)$ on $\gamma$ is controlled by
the nonlinear boundary-value problem \eref{A1}.
Therefore if $\gamma$ is large enough, the second term in \eref{st} may become dominant and so the
anomalous regimes with $dJ/ d \gamma<0$ cannot be ruled out a priori.
Below, we produce several systems that do display this anomalous behaviour.

%

Before proceeding to the analysis of specific examples, it is appropriate to mention that
the  flux from the active to the lossy region bears some analogy with the traffic flow.
Let $\gamma$ be the traffic density on a highway
and $J$  the traffic flow through it.
When the traffic is building up, then,
as long as $J$ stays under the road capacity value,
 the greater the
density, the greater is the traffic flow. However once the capacity of the highway has been reached,
any further increase of $\gamma$ results in the drop of $J$.
The traffic becomes congested.

With this analogy in mind,
 the regime where the inequality $dJ/d\gamma<0$ holds,
will be referred to as the  jamming  anomaly.
(A more detailed discussion of the jamming metaphor appears in section \ref{conclusions}.)

\section{Linear jamming with the double-delta well potential}
\label{double_lin}

Our first choice is the $\mathcal{PT}$-symmetric double-delta well potential
\be
\hspace*{-12mm}
V(x) = - \left[\delta \left(x-\frac{L}{2} \right)  + \delta \left(x+\frac{L}{2} \right)\right] +i \gamma \left[ \delta \left( x+\frac{L}{2} \right)  - \delta \left( x-\frac{L}{2} \right) \right],
\label{C9}
\ee
where $L>0$ is the distance between the wells.
The well on the left (the one at $x=-L/2$) gains
and  the one on the right ($x=L/2$) loses power.

The linear ($g=0$) Schr\"odinger equation \eref{A1} with the potential \eref{C9} has  one or two  discrete
eigenvalues.
The corresponding eigenfunctions are given by the same expression \cite{Uncu,Cartarius}:
\be
\psi(x)= \left\{
\begin{array}{ll}
   \frac{e^{i \phi} + e^{\kappa L- i \phi}}{2 \sqrt N}  e^{\kappa x}, & x \leq -\frac{L}{2};   \\ \\
   \frac{ \cosh (\kappa x+ i \phi)}{\sqrt{N}}, & -\frac{L}{2}  \leq x \leq \frac{L}{2};  \\ \\
 \frac{e^{-i \phi} + e^{ \kappa L+ i \phi}} {2 \sqrt N}   e^{-\kappa x}, & x \geq \frac{L}{2}.
 \end{array}
 \right.
 \label{C5}
 \ee
Here $\kappa$ is a root of the transcendental equation
\be
e^{- 2 \kappa L}= \frac{\gamma^2+ (2 \kappa-1)^2}{\gamma^2+1},
\label{C6}
\ee
and the constant angle $\phi$ is defined by
\[
e^{2 i \phi} = \frac{2 \kappa -1 + i \gamma}{1- i \gamma} e^{ \kappa L}.
\]

Letting
\[
N= \frac{\cos 2 \phi}{2} \left(L+ \frac{1}{\kappa} \right) + \frac{e^{\kappa L}}{2 \kappa},
\]
the eigenfunction \eref{C5} is normalised to unity:
\be
\int_{-\infty}^\infty |\psi|^2 \, dx=1.
\label{N_Lin}
\ee

The analysis of the secular equation \eref{C6}
can be carried out without any use of computer \cite{BZ}.
The number of eigenvalues and the behaviour of the corresponding branches depend on
the distance between the wells.

\begin{figure}[t]
\begin{center}
\includegraphics[width=1.\linewidth]{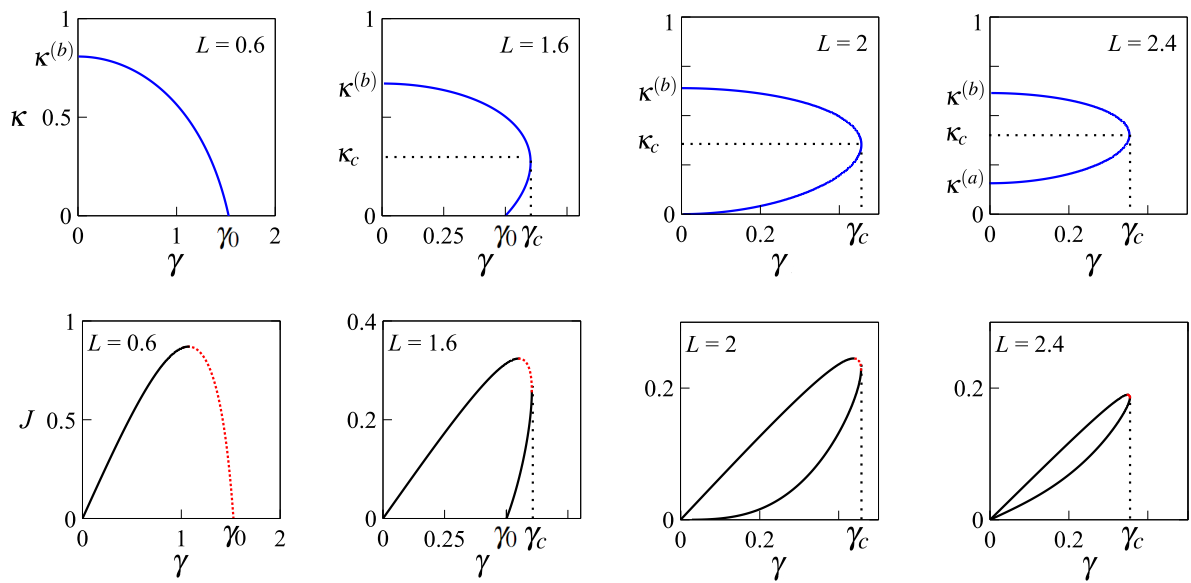}
\end{center}
\caption{\label{8} Positive roots of \eref{C6}  (top row) and the interfacial flux (bottom row)
vs the gain-loss coefficient $\gamma$, for several representative values of $L$.
Here $L<1$ (first column); $1< L <2$ (second column);
$L=2$ (third column), and  $L>2$ (the rightmost column).
 The part of the $J(\gamma)$ curve with the negative slope
 (highlighted in dotted red) represents the jamming anomaly.
}
\end{figure}

When $L<1$, there is a single positive branch of  eigenvalues $\kappa(\gamma)$
which decays, monotonically, as $\gamma$ is increased  from zero to $\gamma_0$
(Fig. \ref{8}, leftmost panel in the top row).
As $\gamma$ reaches $\gamma_0$,
\[
\gamma_0= \sqrt{2/L-1},
\]
the exponent $\kappa$ falls to zero and
the  eigenvalue $-\kappa^2$ immerses in the continuous spectrum.


When $L$ is taken between $1$ and $2$,
then, depending on the interval of $\gamma$ values, the function $\kappa(\gamma)$ has one or two branches
(Fig. \ref{8}, second panel in the top row).
As $\gamma$ is raised from 0 to $\gamma_c$,
one branch descends, monotonically,  from $\kappa^{(b)}$ to $\kappa_c$.
In the interval $\gamma_0< \gamma< \gamma_c$,
there also is a monotonically growing branch.
 On this branch, $\kappa$ increases from 0 to $\kappa_c$ as
$\gamma$ changes  from $\gamma_0$ to $\gamma_c$.
 As $\gamma$ reaches
 $\gamma_c$, the two eigenvalues  merge and become complex.

Finally, when $L \geq 2$, the monotonically increasing and
decreasing  branches of
 $\kappa(\gamma)$ exist over the same interval $0 \leq  \gamma \leq \gamma_c$
 (Fig. \ref{8}, the last two panels in the top row).
As $\gamma$ grows from 0 to $\gamma_c$, one branch of $\kappa$
grows from $\kappa^{(a)}$ to $\kappa_c$ whereas the
other one descends from $\kappa^{(b)}$ to $\kappa_c$.

The interfacial  flux  \eref{A01} associated with the eigenfunction \eref{C5},
 is
\be
J= \frac{\kappa}{N}  \sin 2 \phi.  \label{C60}
\ee
The gain-loss coefficient can be expressed as an explicit function of
the eigenvalue:
\[
\gamma= \sqrt{ \frac{4\kappa(1-\kappa)}{1-e^{-2 \kappa L}} -1}.
\]
Therefore the $N$ and  the $\sin 2 \phi$ factors in \eref{C60}  can also  be explicitly parametrised by $\kappa$.
As a result,  the dependence $J(\gamma)$ can be represented by parametric equations
$\gamma=\gamma(\kappa), J=J(\kappa)$.  This dependence is shown in the panels making up the bottom row in Fig.\ref{8}.
Highlighted are the intervals of $\gamma$ where $dJ/d \gamma<0$.


The variation of $J$ with $\gamma$ is governed by
\[
\frac{d J}{d \gamma}= \frac{dJ / d \kappa}{d \gamma / d \kappa}.
\]
It is instructive to consider the situation where the eigenvalue $-\kappa^2$
approaches the continuous spectrum: $\kappa \to 0$.
In this limit, we have
\[
\frac{dJ}{d \kappa} =  2 \gamma_0+O(\kappa), \quad \frac{d\gamma}{d \kappa}= \frac{L-1}{\gamma_0 L}+O(\kappa).
\]
In case $L<1$, the behaviour of the flux is anomalous: $dJ/ d \gamma <0$. (See Fig.\ref{8}, leftmost panel in the bottom row.)

\section{Nonlinear jamming}
\label{double_nonlin}

To explore jamming in presence of nonlinearity, we  let $g \neq 0$ in
 the stationary equation \eref{A1} with
the same two-delta well potential \eref{C9}. In what follows,
  the inter-well distance  $L$ and the nonlinearity strength $g$ are assumed
  to be fixed.

  Choosing a particular value of $\gamma$ gives rise to a  branch of solutions with a continuously varied $\kappa$,
 rather than a countable set of $\kappa_n$ as in the linear case.
 Since we are interested in the variation of the interfacial flux $J$ due to factors other than  the increase of the
 total power $P=\int |\psi|^2 dx$,
 we  impose the  condition
 \be
 \int |\psi|^2 dx=1.
 \label{E1}
 \ee

 The condition \eref{E1} was imposed in the linear case as well --- recall equation \eref{N_Lin}.
  There, it did not affect the eigenvalues $\kappa$ and was only necessary for the gauging of the flux $J$.
 In contrast, when $g \neq 0$, the condition \eref{E1} selects a particular $\kappa$ from a continuous family and therefore,
  represents a nontrivial constraint.
  The nonlinear Schr\"odinger equation \eref{A1} and this constraint  form a system of two simultaneous equations.
  We note, in passing, that the condition \eref{E1} has a clear physical meaning in the context of the two-trap
  boson-gas interpretation of equation \eref{A1} \cite{Cartarius}.

The system \eref{A1}+\eref{E1} can be solved exactly --- in the sense that its solution can be reduced to a
system of two transcendental equations \cite{BZ}.
Instead  of finding the ``eigenvalue" $\kappa$ and ``eigenfunction" $\psi$ for the given $\gamma$, it
is convenient to assume that the parameter $\kappa$ is given --- and solve for $\psi$ and $\gamma$.
Having  determined $J(\kappa)$ and $\gamma(\kappa)$, one can readily plot $J(\gamma)$ as a parametric curve.

The nonlinear mode is found  \cite{BZ}
 to be
\numparts
\be
\psi(x)=                    \sqrt{ \frac{2}{g} } \kappa  \,       \varphi(x),
\label{E40}
\ee
where
\be
\varphi= \left\{
\begin{array}{lr}
     {\rm sech\/} (\kappa x  + \mu)             e^{-i \chi}, & x \leq -L/2 \\
 \sqrt{\rho (x)}  e^{i \theta(x)}, &  - L/2  \leq x \leq L/2\\
  {\rm sech\/} (\kappa x - \mu) e^{i \chi},
& x \geq L/2
\end{array}
\right..
\label{E20}
\ee
\endnumparts
Here $\chi$ is a  constant phase defined by $\chi= \theta(L/2)$.

In the middle range, $-L/2 \leq x \leq L/2$, the modulus-squared and phase of $\varphi$ are given by
\be
\rho(x)= (\alpha-\beta) {\rm cn}^2 \left( K- \sqrt{2 \alpha+\beta-1} \kappa x, k\right) + \beta
\label{E21}
\ee
and
\[
\theta(x)=  \kappa \sqrt{\alpha \beta (\alpha+\beta-1)} \int_0^{x} \frac{ ds}{\rho(s)},
\]
respectively. Here $\alpha$ and $\beta$ are two constants satisfying
$\alpha \geq \beta \geq 0$ and $\alpha+ \beta >1$.

In equation \eref{E21}, ${\rm cn}$ is the Jacobi elliptic cosine and $K$ is the complete elliptic integral of the first kind.
The elliptic modulus of the Jacobi function ${\rm cn} (z,k)$ and the argument of $K(k)$, is expressible in terms of  $\alpha$ and $\beta$:
\[
k= \sqrt{\frac{\alpha-\beta}{2 \alpha +\beta-1}}.
\]
The parameter $\mu$ in \eref{E20} is also expressible via  these constants:
\be
{\rm sech\/}^2 \left(\mu- \frac{\kappa L}{2} \right)=
\beta+ (\alpha-\beta) \mathrm{cn}^2\left( K-  y,k \right).
 \label{E23}
\ee
Here we have denoted
\[
 y = \frac{L \kappa}{2}
  \sqrt{2 \alpha + \beta -1}.
  \]

The constants $\alpha$ and $\beta$  are determined as two components of a vector root of the
following system of two transcendental equations:
\numparts
\begin{eqnarray}
\zeta^2 +  \beta +(\alpha-\beta) \mathrm{cn}^2   \left( K-  y,k \right) -1=0,
  \label{E24} \\
  \left(  1 + \kappa \zeta  \right)^2 -\kappa^2 S^2=0.
  \label{E25}
\end{eqnarray}
\endnumparts
In \eref{E24}-\eref{E25}, $S^2$ and $\zeta$ are functions of $\alpha$ and $\beta$:
 \begin{eqnarray*}
S^2=
 \frac{  (\alpha+\beta) (\alpha +\beta-1)-
 \alpha \beta }{  \beta+ (\alpha-\beta) \mathrm{cn}^2\left( K-y,k \right)   }
  - \frac{\alpha \beta (\alpha+\beta  -1)}{\left[    \beta+ (\alpha-\beta) \mathrm{cn}^2\left( K-y,k  \right)           \right]^2}
  \nonumber    \\
    \phantom{\frac{\frac12}{\sqrt{A^2}}}
  +1
 -\beta-  (\alpha-\beta) \mathrm{cn}^2\left( K-y,k  \right)
\end{eqnarray*}
and
\[
\hspace*{-7mm}
\zeta= \frac{\alpha+\beta-1}{\sqrt{2 \alpha+\beta-1}} y -1  +\frac{g}{4 \kappa}  -
\sqrt{2 \alpha+\beta-1}
\left\{ E \left( \frac{\pi}{2},k \right) -
 E \left[ \mathrm{am} \left( K- y \right), k  \right] \right\},
 \]
where $E[\mathrm{am}(z),k]$ is the incomplete elliptic integral of the second kind,
\[
E[\mathrm{am}(z),k]= \int_0^{\mathrm{am} (z)} \sqrt{1-k^2 \sin^2 \theta} \, d \theta =
\int_0^z \mathrm{dn}^2 (w,k) dw,
\]
and  $E \left( \frac{\pi}{2},k \right)$ is the complete  elliptic integral:
\[
E \left( \frac{\pi}{2},k \right)=  \int_0^{\pi/2} \sqrt{1-k^2 \sin^2 \theta} \,  d \theta.
\]

\begin{figure}[t]
\begin{center}
 \includegraphics[width=0.7\linewidth]{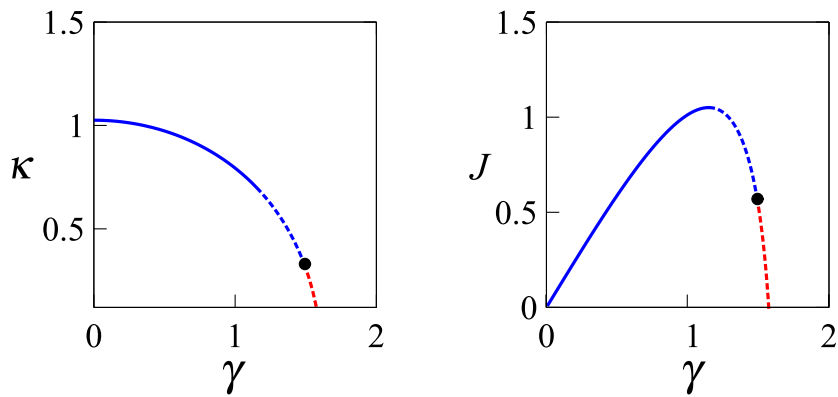}
\end{center}
\caption{\label{nl}
The ``nonlinear eigenvalue" $\kappa$ (left) and the
interfacial flux $J$ (right) as a function of the gain-loss coefficient $\gamma$ in the nonlinear Schr\"odinger equation
with the double-delta well potential \eref{C9}.
In this plot, $g=1$ and $L=0.6$.
The interval where $dJ/d\gamma<0$ pertains to  the jamming anomaly (shown by the dashed  curves).
The blue and red parts of the jammed curves indicate stable and unstable regimes, respectively.}
\end{figure}

Note that the transcendental equations \eref{E24}-\eref{E25} involve an
explicit dependence on $\kappa$ but not on $\gamma$. Hence
$\alpha=\alpha(\kappa)$, $\beta=\beta(\kappa)$.
Once a root of \eref{E24}-\eref{E25} has been determined,
the gain-loss coefficient is expressible as
\be
\gamma(\kappa)=
\frac
{\sqrt{\alpha \beta (\alpha+\beta-1)} \kappa}
{
\beta+ (\alpha-\beta) \mathrm{cn}^2\left( K-y, k \right)
}.
\label{gCD}
\ee
Substituting the function \eref{E40}-\eref{E20} in \eref{D20} gives the interfacial flux $J$, also as
a function of $\kappa$:
\be
J(\kappa)= 2 \kappa \sqrt{\alpha \beta (\alpha+\beta-1)}.
\ee

Solving the transcendental system \eref{E24}-\eref{E25} for a range of $\kappa$-values,
we  obtain the $J(\gamma)$ dependence in the parametric form: $\gamma=\gamma(\kappa)$,
$J=J(\kappa)$.
This is plotted in Fig.\ref{nl}, for a  representative pair of $L$ and $g$.
The dependence of the flux on the gain-loss coefficient features a segment with a negative slope
indicating the jamming anomaly.

The vital property of a nonlinear mode is its stability.
To classify the stability of the stationary solution \eref{E40}-\eref{E20}, we consider
a solution to
equation \eref{A0} of the form
\[
\Psi(x,t)= e^{i \kappa^2 t}\left\{  \psi(x) +\epsilon  \, e^{\lambda t } \left[ p(x)+ iq(x)\right] \right\},
\]
with $p$ and $q$  real.
Linearising in small $\epsilon$
 gives an eigenvalue problem
\be
\mathcal H {\vec p}=  \lambda J {\vec p},
\label{EV}
\ee
where $\mathcal H$
and $J$   are $2 \times 2$ matrices, and
${\vec p}$ is a two-component vector-column:
\be
\mathcal H= - I \frac{d^2}{dx^2} + \left( \begin{array}{lr} \kappa^2+ U & -W \\
W & \kappa^2+ U \end{array} \right)
-g \left( \begin{array}{lr} 3 {\mathcal R}^2 + {\mathcal I}^2 &  - 2{\mathcal R}{\mathcal I} \\
2 {\mathcal R}{\mathcal I}  & 3 {\mathcal I}^2 + {\mathcal R}^2 \end{array}
\right),
\label{E50}
\ee
\[
J= \left( \begin{array}{lr} 0 & -1 \\
1 & 0
\end{array} \right),
\quad
I= \left( \begin{array}{lr} 1 & 0 \\
0 & 1
\end{array} \right),
\quad
{\vec p}= \left( \begin{array}{c}  p \\ q \end{array} \right).
\]
In \eref{E50}, $\mathcal R$ and $\mathcal I$ are the real and imaginary part of  $\psi(x)$:
\[
\psi= \mathcal R(x) + i \mathcal I(x).
\]
The stationary solution $\Psi= e^{i \kappa^2 t} \psi(x)$ is classified as unstable if the operator $J^{-1} \mathcal H$ has  (at least one) eigenvalue
$\lambda$ with a positive real part.

The eigenvalue problem \eref{EV} was solved numerically.
The resulting stability and instability domains are demarcated in Fig.\ref{nl}.
The central conclusion of this computer analysis is that the stationary nonlinear mode  \eref{E40}-\eref{E20} is stable in a
sizeable interval of the gain-loss amplitudes.

\section{Jamming with the \PT-symmetric Scarff potential}
\label{Sca}

Does one need to have {\it two\/} potential wells in order to observe the jamming effect?
The aim of this section is to demonstrate the same phenomenon in a {\it single-}well
potential --- yet with the bicentric distribution of gain and loss.

This complex profile is known as the \PT-symmetric Scarff II potential \cite{BQ1,BQ2,Ahmed}:
\be
V(x)= -q \, {\rm sech}^2 \, x - i \gamma  \, {\rm sech} \, x \tanh x.
\label{D1}
\ee
This time the potential well is centred at the origin, while the gain and loss are continuously
distributed over the entire left (gain) and right (loss) semiaxis.

The sufficiently deep potential well \eref{D1} can support an arbitrarily large number of bound states.
For simplicity, we focus on the two lowest eigenvalues pertaining to the single-hump eigenfunctions.
By the direct substitution one can verify that the
 eigenvalues  are given by
\begin{eqnarray}
\kappa_1= \frac{ \sqrt{\gamma_c+\gamma} +\sqrt{\gamma_c-\gamma}}{2}-\frac12, \nonumber \\
\kappa_2=  \frac{ \sqrt{\gamma_c+\gamma} -\sqrt{\gamma_c-\gamma}}{2}-\frac12, 
\nonumber
\end{eqnarray}
where
\[
\gamma_c= q+\frac14,
\]
and we have assumed that $\gamma  \leq \gamma_c$.
The corresponding eigenfunctions are
\begin{eqnarray}
 \psi_1=  \pi^{-1/4}   \sqrt{  \frac{\Gamma(\kappa_1+ 1/2)}{\Gamma(\kappa_1)} }
 ({\rm sech} \, x)^{\kappa_1}  \exp \left\{   i \frac{2 \kappa_2+1}{2}  \arctan (\sinh x) \right\},
 \nonumber
 \\
  \psi_2=  \pi^{-1/4} \sqrt{\frac{\Gamma(\kappa_2+1/2)}{\Gamma(\kappa_2)} }
 ({\rm sech} \, x)^{\kappa_2}  \exp \left\{   i \frac{2 \kappa_1+1}{2}  \arctan (\sinh x) \right\},
 \label{D3}
\end{eqnarray}
where $\Gamma(s)$ is the gamma function.
(For the complete list  of eigenvalues and eigenfunctions, see \cite{Ahmed}.)

\begin{figure}[t]
\begin{center}
 \includegraphics[width=85mm]{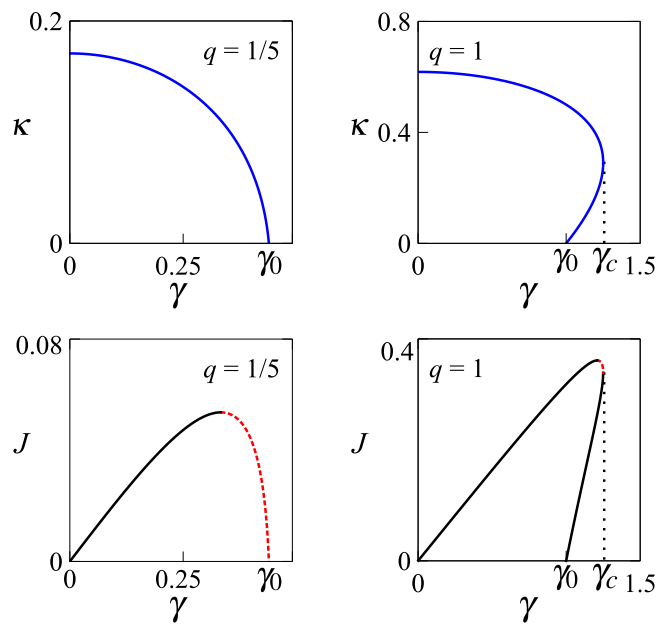}
\end{center}
\caption{\label{Scarff}
The eigenvalues (top row) and interfacial flux (bottom row)  for the \PT-symmetric Scarff II potential.
The left column pertains to $q<\frac14$ and the right one  to $q> \frac14$.
 In the bottom row, the part of the curve with the negative slope
 (highlighted in dotted red) represents the jamming anomaly.
}
\end{figure}

When $q< \frac14$, there is a single eigenvalue $\kappa_1$.
As $\gamma$ grows from 0 to $\gamma_0$, where $\gamma_0= \sqrt q$,  the branch $\kappa_1$ descends, monotonically,
from $\sqrt{\gamma_c}-\frac12>0$ to zero, where the eigenvalue immerses in the continuous spectrum.

When $q> \frac14$,  we have two branches. As $\gamma$ changes from zero to $\gamma_c$,
the branch $\kappa_1$ descends from $\sqrt{\gamma_c}-1/2$ to $\kappa_c$, where $\kappa_c=\frac12 (\sqrt{2 \gamma_c}-1)>0$.
In the interval $(\gamma_0, \gamma_c)$, there also is a monotonically growing branch, $\kappa_2(\gamma)$.
As $\gamma$ varies from $\gamma_0$ to $\gamma_c$, the eigenvalue $\kappa_2$ grows from zero
and  merges with $\kappa_1$.

Substituting \eref{D3} into \eref{D20} gives the flux across the gain-loss interface,
associated with the two eigenfunctions:
\be
J_1(\gamma)=   \frac{ 2 \kappa_2+1}{\sqrt{\pi}}  \frac{\Gamma( \kappa_1+\frac12)}{\Gamma(\kappa_1)},
\quad
J_2(\gamma)=   \frac{ 2 \kappa_1+1}{\sqrt{\pi}}     \frac{\Gamma( \kappa_2+\frac12 )}{\Gamma (\kappa_2)}.
\label{D4}
\ee
The dependencies \eref{D4} are shown in the bottom row of Fig.\ref{Scarff}.
In particular, when $q<\frac14$, we have
\[
\frac{dJ_1}{d \gamma} \to  \frac{q}{q-1/4}<0  \ \mbox{as} \ \kappa_1 \to 0.
\]
As in the double-delta potential
(Fig.\ref{8}), the behaviour of the flux is anomalous near the point $\gamma=\gamma_0$
where the eigenvalue immerses in the continuous spectrum.

\section{The flux anomaly near the exceptional point}
\label{exceptional}

The double-delta well and the Scarff potential exhibit a similar behaviour of the flux
near the exceptional point; namely, one branch of eigenvalues has $J(\gamma)$ approaching
$J(\gamma_c)$ at the infinite negative and the other one at the infinite positive slope.
Our objective here is to show that this square-root behaviour is common for all \PT-symmetric potentials with
exceptional points. (In particular, this implies that an exceptional point cannot be a cusp of $J(\gamma)$.)

Consider an eigenvalue problem
\be
H \psi=E \psi,
\label{D7}
\ee
for the
 \PT-symmetric Schr\"odinger operator
\be
H= -\frac{d^2}{dx^2} + U(x) +i \gamma \mathcal{W}(x),
\label{D14}
\ee
where
 the functions $U(x)$ and  $\mathcal{W}(x)$ are real, decay to zero as $|x| \to \infty$
and satisfy $U(-x)=U(x)$ and $\mathcal W(-x)=- \mathcal W(x)$.

Let
the
operator \eref{D14}  have two real eigenvalues, $E^{(a)}$ and $E^{(b)}$, $E^{(a,b)}<0$,
for all $\gamma< \gamma_c$  in some neighbourhood of $\gamma_c$, and
denote  the corresponding  eigenfunctions $\psi^{(a)}(x)$ and $\psi^{(b)}(x)$.
Assume that  the two eigenfunctions coalesce as $\gamma$ approaches  $\gamma_c$:
\[
\psi^{(a)}, \psi^{(b)} \to \psi_0,
\quad
E^{(a)}, E^{(b)} \to E_0
\quad \mbox{as} \ \gamma \to \gamma_c.
\]
 Therefore, the value $\gamma=\gamma_c$ is
 an exceptional point of the operator $H$,
while $E_0$ is a repeated  eigenvalue of algebraic multiplicity two:
\begin{eqnarray}
H_0 \psi_0= E_0 \psi_0,        \label{D11}   \\
(H_0-E_0) \phi_0=\psi_0.
\label{D10}
\end{eqnarray}
Here
\[
H_0= -\frac{d^2}{dx^2} + U(x) +i \gamma_c \mathcal{W}(x);
\]
 $\psi_0(x)$ is the eigenvector associated with the eigenvalue $E_0$ and
$\phi_0(x)$ is the generalised eigenvector.

Without loss of generality we can choose $\psi_0$ to be \PT-symmetric:
\[
\psi_0^*(-x)=\psi_0(x).
\]
(Indeed, if $\psi(x)$ is an eigenvector associated with a real eigenvalue $E$, then so is the sum $\psi(x)+\psi^*(-x)$. This sum defines a \PT-symmetric function.)
Furthermore, if $\phi_0(x)$ is a solution to equation \eref{D10} with a \PT-symmetric right-hand side $\psi_0$, then so is $\phi_0(x)+\phi_0^*(-x)$.
Therefore $\phi_0$ can also be chosen to be \PT-symmetric:
\[
\phi_0^*(-x)=\phi_0(x).
\]

The \PT-symmetric solution $\phi_0$ of equation \eref{D10} is defined up to the addition of an arbitrary real multiple of $\psi_0$.
It is convenient to choose this multiple in such a way that
\be
\int \psi_0^* \phi_0  \, dx=0.
\label{D32}
\ee
We assume $\psi_0$ to be normalised to unity,
\be
\int \psi_0^*  \psi_0  dx=1,
\label{D30}
\ee
and note the following identity which follows from
the  equation \eref{D10}:
\be
\int  \psi_0^2 \, dx=0.
\label{D8}
\ee

Letting
\be
 \epsilon^2 = \gamma_c-\gamma >0,
\label{D25}
\ee
we  expand the eigenvalue and eigenfunction in powers of $\epsilon$:
\begin{eqnarray}
E= E_0+ \epsilon E_1 + \epsilon^2 E_2+ ..., \nonumber \\
 \psi(x) =\psi_0(x) + \epsilon \psi_1(x) + \epsilon^2 \psi_2(x)+ ... .
 \label{D22}
\end{eqnarray}
Substituting these expansions
in \eref{D7}, we equate coefficients of like powers of $\epsilon$.
At the order $\epsilon^1$, we have
\[
(H_0-E_0) \psi_1= E_1 \psi_0.
\]
Using equation \eref{D10} we obtain a solution:
\be
\psi_1(x)= E_1 \phi_0(x)  +\eta \psi_0(x),
\label{D12}
\ee
where $\eta$ is an arbitrary constant coefficient.
In order to ensure that $\psi_1$ is \PT-symmetric, this coefficient has to be real.

The order $\epsilon^2$ gives
\[
(H_0-E_0) \psi_2 - i \mathcal W \psi_0 = E_2 \psi_0+ E_1 \psi_1.
\]
The solvability condition for $\psi_2$ gives
\be
E_1^2= \frac{1}{i} \frac{\int  \mathcal W \psi_0^2 \,  dx}{ \int \psi_0 \phi_0  \, dx},
\label{D18}
\ee
where we have made use of  \eref{D8}  and \eref{D12}.

Using the symmetry of $\psi_0(x)$, $\phi_0(x)$ and $\mathcal W(x)$,
one can readily check that
the quotient \eref{D18} is real.
The energies $E^{(a)}$ and $E^{(b)}$ are given by $E_0+ \epsilon E_1$
and $E_0-\epsilon E_1$,  respectively, where $E_1$ is one of the two opposite roots in \eref{D18}.
By assumption, $E^{(a)}$ and $E^{(b)}$ are real; hence the quotient \eref{D18}
should be positive. (The negative quotient \eref{D18} would  simply imply
that the real eigenvalues $E^{(a,b)}$ pertain to the negative, not positive, parameter $\epsilon^2$ in \eref{D25}.)

For the purposes of the flux analysis, it is essential that the eigenfunction $\psi(x)$ be normalised to unity:
\be
\int \psi^* \psi \,  dx=1.
\label{D40}
\ee
Substituting from \eref{D22} gives
\[
\int \psi^* \psi \,  dx= 1+ 2  \eta \epsilon +O(\epsilon^2),
\]
where we have used the identity \eref{D32}. Thus if we choose $\eta=0$, we will ensure the normalisation condition \eref{D40}
 to order $\epsilon$.

Substituting the expansion of the eigenfunction \eref{D22} in
 \eref{D20} we obtain the dependence of the interfacial flux on $\gamma$
 in the vicinity of the exceptional point:
\[
J^{(a,b)}=J_0 \pm  (\gamma_c-\gamma)^{1/2}  J_1 +  O(\gamma_c-\gamma),
\]
where
\begin{eqnarray}
J_0= \left.
 i \left(  \frac{d \psi_0^* }{dx} \psi_0- \psi_0^*  \frac{d \psi_0}{d x} \right) \right|_{x=0},
 \nonumber \\
 J_1=i E_1 \left[ \left( \frac{d\phi_0^*}{dx} \psi_0 + \frac{d \psi_0^*}{dx} \phi_0 \right) - c.c.
 \right]_{x=0}.
 \label{D23}
 \end{eqnarray}
 (In \eref{D23}, the {\it c.c.} stands for the complex conjugate of the preceding term.)

 The curves $J^{(a)}(\gamma)$ and $J^{(b)}(\gamma)$ have opposite slopes in the
 vicinity of $\gamma_c$:
 \[
 \frac{d J^{(a)}}{d \gamma} = -\frac{J_1}{2 \sqrt{\gamma_c-\gamma}} + O(1),
 \quad
  \frac{d J^{(b)}}{d \gamma} = \frac{J_1}{2 \sqrt{\gamma_c-\gamma}} + O(1).
  \]
 One of the two dependencies is anomalous, $dJ/d\gamma<0$.
 This is exemplified by Fig.\ref{8} (second, third and forth panels in the bottom row) and Fig.\ref{Scarff} (bottom right panel).

The upshot of this analysis is that there always is a jammed branch near the exceptional point.

\section{No jamming in the \PT-symmetric  parabolic   potential}
\label{harmonic}

Does every \PT-symmetric potential exhibit an interval of the gain and loss amplitude
with the anomalous behaviour of the interfacial flux?
In this section we produce an exactly solvable counter-example.

More specifically, we consider a \PT-symmetric harmonic oscillator of the form
\be
V(x)= x^2 - 2 i \gamma x, \quad \gamma>0. \label{ho}
\ee
The Schr\"odinger operator \eref{A1} with this potential supports an  infinite sequence of real eigenvalues $-\kappa^2_n= 2n+1+\gamma^2$,  where $n=0,1,2, ...$
\cite{parab1,parab2}.
The corresponding $L^2$-normalized eigenvectors are given by \cite{parab3}
\begin{equation}
\psi_n(x)=   \frac{\pi^{-1/4}}{\sqrt{ 2^n n!}}
 \frac{e^{-\gamma^2/2}}{  \sqrt{ L_n(-2\gamma^2)}} H_n(x-i\gamma)e^{-(x-i\gamma)^2/2}.
 \label{D5}
\end{equation}
In \eref{D5},
 $H_n(x)$ and $L_n(x)$ are the $n$-th order Hermite and Laguerre polynomials, respectively.

Substituting \eref{D5} in \eref{D20} one can determine the value of the flux associated with each eigenvector.
The first three values are given  by
\begin{eqnarray}
J_0(\gamma) =\frac{1}{\sqrt{\pi}} \, \gamma,   \quad
J_1(\gamma) =  \frac{2}{\sqrt{\pi}}  \,   \frac{\gamma(\gamma^2+1)}{ 2\gamma^2+1},
\nonumber \\
J_2(\gamma) = \frac{1}{2 \sqrt{\pi}} \,
\frac{\gamma(4\gamma^4+12\gamma^2+5)}{2\gamma^4+4\gamma^2+1}.
\label{D6}
\end{eqnarray}
As one can readily check, each of the  three expressions \eref{D6} defines a monotonically
 growing function of $\gamma$. Hence no jamming behaviour is exhibited by the \PT-symmetric harmonic oscillator.

This conclusion  is consistent with results of the previous section where we have identified exceptional points as one source of the jamming anomalies.
In the system at hand, the \PT-symmetry remains unbroken for an arbitrarily large $\gamma$ and the spectrum does not feature any exceptional points.

Eigenvalue immersions in the continuous spectrum were also seen to be accompanied by jamming (Figs. \ref{8}, \ref{nl}, and \ref{Scarff}).
The operator \eref{A1} with the parabolic potential \eref{ho} does not have  any continuous spectrum; hence this source of anomalous
behaviour was not available to the oscillator either.

\section{Concluding remarks}
\label{conclusions}

The purpose of this paper was to describe an anomalous
phenomenon occurring in \PT-symmetric systems
of optics and atomic physics. The   counter-intuitive
effect    consists in the reduction of the power gain (or the particle influx) in the active
part of the system accompanied by the reduction of the power loss (or particle leakage) in its dissipative part,
as a result of of the increase of  the gain-loss coefficient. The reduction of the power gain and loss in their respective parts of the
system manifests itself in the drop of the  flux between the two parts.

We have been referring to this phenomenon as {\it jamming}, because of the analogy it bears
to the behaviour of   the traffic flow
through a road network.  Here, we explain the traffic analogy in some more detail.

\subsection{Jamming in a parallel road network}

Consider the centre of a city connected to its large suburb
or a satellite town by $N$ alternative routes $R_1, R_2, ..., R_N$.
 Let $\mathcal W_n$ stand for the average free-flow speed on the  road $R_n$ of this network.
Typically, the distribution of velocities  $\mathcal W_n$ will have a single maximum, say $\mathcal W_1$, pertaining to a  highway or  toll road.
 Roads with traffic lights, $R_2, ..., R_\ell$,  will offer lower characteristic speeds, and  routes $R_{\ell+1}, ..., R_N$  through the residential areas will have
even smaller values of $\mathcal W_n$ due to severe speed limits.

The traffic flow on the road $R_n$
is $\mathcal W_n \rho_n$, where  $\rho_n$ is the  corresponding traffic density.  The total traffic flow is $J= \sum_{n=1}^N  \mathcal W_n \rho_n$.

It is convenient to think of  the density $\rho_n$  as a product
 $\gamma |\psi_n|^2$, where $\gamma$ is a factor accounting for the diurnal density variations
 and
 $|\psi_n|^2$  is the probability to find a  car entering the network, at the entrance to the  road $R_n$.
Note that $\sum_{n=1}^N |\psi_n|^2=1$.

When the traffic is low ($\gamma$ small),
 the distribution $|\psi_n|^2$ depends on  $\gamma$ only weakly and has a maximum at $n=1$.
 However as $\gamma$ grows and the flow $\gamma \mathcal W_1 |\psi_1|^2$ exceeds the carrying capacity of the highway $R_1$,
  the distribution starts changing
 --- the maximum value $|\psi_1|^2$ starts decreasing and the density leaks to ``slower" roads, $R_2, ..., R_\ell$.
 The vehicles opt for the secondary routes  in order to avoid the congestion
 on the highway.

 As $\gamma$ keeps growing, some other carrying capacities are exceeded and the density distribution $|\psi_n|^2$
 flattens further.  (Some desperate motorists try to make their way through the  streets  $R_{\ell+1}, ..., R_N$.)
 If the density spreading proceeds faster than the growth of $\gamma$,
the total traffic flow $J(\gamma)$ will reach a maximum and start decreasing --- despite
the growth of the number of vehicles in the network.
This is a metaphor of the phenomenon we have detected in several \PT-symmetric systems.

\subsection{Summary and concluding remarks}

We have demonstrated that the jamming anomaly in the
\PT-symmetric linear Schr\"odinger equation
 may be anticipated in two broad classes of situations.
First, the jamming occurs in the vicinity of the exceptional point,
 where
two real eigenvalues coalesce and acquire imaginary parts.
This general conclusion was exemplified by the double-delta well potential with a large distance between the wells
($L>1$) and the deep single-well Scarff potential  ($q>1/4$).
In contrast, the \PT-symmetric harmonic oscillator --- the potential that does not support any exceptional points ---
was shown to be anomaly-free.

Second, the flux associated with an eigenfunction $\psi(x)$ decreases as the corresponding eigenvalue
approaches the edge of the continuous spectrum.
This law admits a simple explanation.
As $\kappa \to 0$, the effective width of the
corresponding eigenfunction grows
as $1/\kappa$. Since the $L^2$-norm of the eigenfunction $\psi(x)$  remains equal to 1,
this implies that $|\psi(0)|$, the maximum value of the single-hump eigenfunction, tends to zero
(in proportion to $\kappa^{1/2}$).
As a result, the flux decreases: $J(\gamma) -J(\gamma_0) \sim \kappa$.

Thus, if it is the {\it increase\/} of $\gamma$ that drives the eigenvalue to the continuum,
 the slope of the $J(\gamma)$ curve
near the immersion point will be anomalous.
The examples are the double-delta well potential with $L<1$ and the Scarff potential with $q<1/4$.

We have also demonstrated that the jamming anomaly can occur in  nonlinear systems.
Specifically, we computed a branch of nonlinear modes supported by the  \PT-symmetric double-delta potential
and observed the nonmonotonic dependence of the flux on the gain-loss amplitude.
It is worth noting that the nonlinear modes exhibiting the anomalous behaviour can be dynamically stable.
The corresponding physical regimes are robust and should be experimentally detectable.

We conclude this section with two remarks.
The first one is on
the similarity and differences between the jamming anomaly and  the macroscopic Zeno effect.

 \textcolor{black}{The {\it macroscopic\/} Zeno effect is the descendant of its celebrated {\it quantum\/} namesake \cite{quantum_Zeno}. It consists in the drop of the external current employed to compensate  losses in the boson condensate ---
  as a result of
 boosting the damping rate at some local sites   \cite{Zeno-theor1,ZKBO}.
 Experimentally,  the  effect is
 manifested in the  reduction of the decay rate of atoms as the  strength of the localised dissipation is increased \cite{Ott}.
 Similar to the
jamming anomaly, an increase of the dissipation coefficient produces a drop in losses
here. However, there are  notable differences as well. In particular,
 raising the damping coefficient in the \PT-symmetric system requires the simultaneous
 and symmetric increase of its gain factor.}

\textcolor{black}{
Since the light propagating in coupled  waveguides obeys the same system of damped  nonlinear Schr\"odinger equations as the boson condensate loaded in the double-well trap,
the macroscopic Zeno effect  has an optical analogue \cite{Zeno-optics}.
Its essential feature is  the suppression of
the light absorption in the waveguide coupler as the dissipation coefficient  in one of its arms exceeds a critical
value.
A closely related phenomenon was observed in coupled silica micro-toroid resonators
 \cite{Peng_diss}. When the parameters of the two-resonator structure were chosen in the vicinity of its spectral exceptional point, the threshold of the Raman
 lasing was seen to be lowered by raising the dissipation coefficient.
 }

\textcolor{black}{
The macroscopic Zeno effect, in condensates and in optics,  as well as the enhancement of lasing in microresonators \cite{Peng_diss}, are
similar to the jamming anomaly in that the increase of the dissipation coefficient leads to the decrease of losses.
One dissimilarity  between the condensate traps  \cite{Zeno-theor1,ZKBO} and optical  couplers \cite{Zeno-optics,Peng_diss} on the one hand,
and our \PT-symmetric system on the other ---
is that the former are purely dissipative set-ups that rely upon the energy or particle influx  ``from infinity" (i.e., from outside). In contrast,  the \PT-symmetric system
balances its losses with an internal gain.
Jamming  is an internal property of the bicentric gain-loss configuration stemming from its short-range structure.
}

\textcolor{black}{
 Another difference is that the Zeno effect and the lasing enhancement can be understood using just two modes
whereas the jamming anomaly is a collective phenomenon which requires both active and lossy subsystems to
have an internal structure. There is no jamming in the two-mode system; see section \ref{toy}.
}

As the second remark, we note that
the jamming anomaly provides a simple checking mechanism for the energy captured in the bicentric structure.
Consider, for instance, the double-well potential \eref{C9}
and assume $L<1$.
The quantities
$|\psi(\pm L/2)|^2$ give the intensity of light in the waveguides with loss and gain, respectively.
Using equation \eref{A2}, these are expressible as
\[
 |\psi( \pm L/2)|^2 = \frac{J(\gamma)}{2 \gamma}.
\]
In the range of gain and loss  where $dJ/d \gamma \leq 0$, that is, in the range $\gamma_* \leq  \gamma < \gamma_0$,
the intensity is bounded by
its value at the point where $J(\gamma)$ is maximum:
\[
|\psi( \pm L/2)|^2 \leq  \frac{J(\gamma_*)}{2 \gamma_*}.
\]

\ack
We thank Nora Alexeeva for computational assistance and Boris Malomed for useful discussions.
This work was funded by the NRF of South Africa (grants UID 85751, 86991, and 87814) and the FCT of Portugal through the grant  PTDC/FIS-OPT/1918/2012.
The financial support from the UCT Open Access Journal Publications Fund
is also gratefully acknowledged.

\end{document}